\documentclass[twocolumn,showpacs,preprintnumbers,amsmath,amssymb,reprint,prl,aps,superscriptaddress]{revtex4-1}
\usepackage{graphicx}
\usepackage{amsmath}
\usepackage{verbatim}
\usepackage{hyperref}

\newcommand{\dBm}{\mathrm{dBm}}

\newcommand{\ns}{\mathrm{ns}}

\newcommand{\dB}{\mathrm{dB}}

\newcommand{\ket}[1]{\left\lvert #1 \right\rangle}
\newcommand{\bra}[1]{\left\langle #1 \right\rvert}

\newcommand{\MP}{\mathcal{M}_{\mathrm{P1,2}}}

\newcommand{\Mp}{\mathcal{M}_{\mathrm{p}}}

\newcommand{\tauMA}{\tau_{\mathrm{MA}}}
\newcommand{\tMD}{t_{\mathrm{MD}}}
\renewcommand{\MP}{M_{\mathrm{P}}}
\newcommand{\unit}[1]{\mathrm{#1}}
\newcommand{\VHA}{V_{\mathrm{H},A}}

\newcommand{\QAn}{A}
\newcommand{\QDi}{D_1}
\newcommand{\QDii}{D_2}

\newcommand{\gDi}{g_{1}}
\newcommand{\eDi}{e_{1}}
\newcommand{\fDi}{f_{1}}

\newcommand{\gAn}{g_{A}}
\newcommand{\eAn}{e_{A}}

\newcommand{\gDii}{g_{2}}
\newcommand{\eDii}{e_{2}}

\newcommand{\txt}[1]{\mathrm{#1}}

\newcommand{\Fa}{\mathcal{F}_{\mathrm{A}}}

\newcommand{\iswap}{$i$\textsc{Swap}}
\newcommand{\cphase}{\textsc{c-Phase}}

\begin{document}
\title{Entanglement genesis by ancilla-based parity measurement in 2D circuit QED}
\author{O.-P.~Saira}
\author{J.P.~Groen}
\author{J.~Cramer}
\author{M.~Meretska}
\author{G.~de~Lange}
\author{L.~DiCarlo}
\affiliation{Kavli Institute of Nanoscience, Delft University of Technology, P.O. Box 5046,
2600 GA Delft, The Netherlands}
\date{\today}

\begin{abstract}
We present an indirect two-qubit parity meter in planar circuit quantum electrodynamics, realized by discrete interaction with an ancilla and a subsequent projective ancilla measurement with a dedicated, dispersively coupled resonator. Quantum process tomography and successful entanglement by measurement demonstrate that the meter is intrinsically quantum non-demolition. Separate interaction and measurement steps allow commencing subsequent data qubit operations in parallel with ancilla measurement, offering time savings over continuous schemes.
\end{abstract}

\maketitle

Controlling the entanglement between qubits is central to the development of every quantum computing architecture. Early efforts with superconducting quantum circuits relied on quantum interference for this purpose. Programmed sequences of one- and few-qubit gates fitting within qubit coherence times have allowed the generation of two- and three-qubit entanglement~\cite{Steffen06,Ansmann09,Dicarlo10,Neeley10}, and the implementation of elementary quantum algorithms~\cite{DiCarlo09,Yamamoto10,Reed12,Dewes12,Lucero12} and games~\cite{Mariantoni11b}.

Recently, focus has shifted toward generating and preserving entanglement by non-demolition measurement of multi-qubit observables, and their use in feedback loops as required for quantum error correction~\cite{Devoret13}. Of particular interest is the parity measurement~\cite{Ruskov03,Lalumiere10,*Tornberg10} that discriminates between states in a multi-qubit register with even or odd total excitation number. Parity measurement on four data qubits at the corners of every square tile on a lattice is needed to realize surface codes, offering the highest fault-tolerance thresholds to date~\cite{Bravyi98,*Raussendorf07,Fowler12}.

A convenient approach to implementing a parity measurement is a two-step indirect scheme involving coherent interaction of the data qubits with an ancillary qubit and subsequent strong measurement of this ancilla. To date, indirect four-qubit parity measurements have been achieved only in trapped-ion systems~\cite{Barreiro11}. In the solid state, parity measurement using an ancillary electron spin has been used to generate probabilistic entanglement between two nuclear spins in nitrogen-vacancy centers in diamond~\cite{Pfaff12}. More recently, parity measurement of two transmon qubits using a dispersively coupled 3D cavity has been used in a digital feedback loop to generate entanglement deterministically~\cite{Riste13c}. An important next step is the realization of parity measurements in an architecture amenable to surface coding.

In this Letter, we present an ancilla-based two-qubit parity measurement in a planar circuit QED architecture~\cite{Wallraff04}. Tomographic characterization shows that dephasing within even and odd parity subspaces is due to intrinsic qubit decoherence during interaction and measurement steps, making the parity meter intrinsically quantum non-demolition (QND). As a further demonstration of this non-demolition character, we generate entanglement by parity measurement on a maximal superposition state. Performing all tomographic data-qubit operations after the ancilla measurement, we achieve a concurrence of 0.46 (0.38) in the even (odd) measurement outcome using a single threshold for conditioning on the ancilla readout, matching the open-loop performance of the recent implementation based on continuous measurement~\cite{Riste13c}. A distinct architectural advantage of our two-step scheme is the possibility to continue operations on the data qubits while the ancilla measurement is performed. Performing the entanglement-by-measurement protocol using such parallel timing instead, the concurrence in the even (odd) parity outcome improves to 0.74 (0.63).

\begin{figure}
\includegraphics[width=\columnwidth]{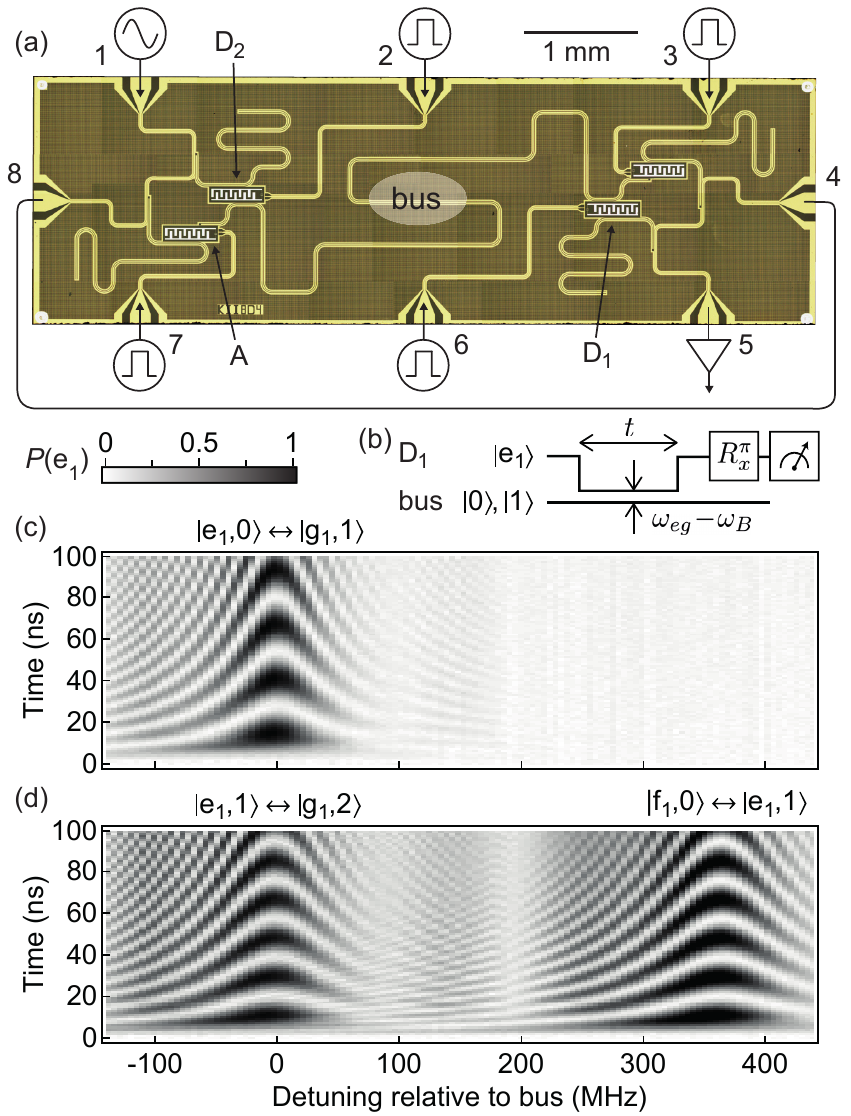}
\caption{(a) cQED processor with four transmons ($\QAn$, $\QDi$, $\QDii$, and unused one at top right) coupled to a bus resonator. Each transmon has a dedicated readout resonator that is addressed through the shared feedline (ports 1--8--4--5), and a local flux-bias line (ports 2, 3, 6, 7) that allows tuning of the transition frequencies with $\sim$1~ns resolution~\cite{DiCarlo09}. A coaxial cable connects ports 4 and 8 off the chip. (b) Gate sequence for coherent swapping of excitations between $\QDi$ and the bus by non-adiabatic qubit tuning. \hbox{(c), (d)} Measured average qubit populations at the end of the sequence for the 1- and 2-excitation manifolds, respectively. An excitation can be swapped from $\QDi$ to the bus (or vice versa) in $13.1\,\unit{ns}$ when the $g_1$--$e_1$ transition is resonant with $B$. In the 2-excitation manifold (d), population transfer occurs when either the $g_1$--$e_1$ or the $e_1$--$f_1$ transition is resonant with $B$ with a half-period of $9.3\,\unit{ns}$. The ratio of the measured periods agrees with the $\sqrt{2}$ enhancement of the effective coupling predicted by theory.
}
\label{Fig1}
\end{figure}

Our quantum processor, shown in Fig.~1(a), combines four transmon qubits (data qubits $\QDi$ and $\QDii$, ancilla $\QAn$, and fourth unused qubit) and five resonators, expanding the architecture introduced in Ref.~\onlinecite{Groen13}. A high-$Q$ resonator bus ($B$) couples to every qubit and mediates all interactions. Dedicated resonators, each dispersively coupled to one qubit, allow frequency-multiplexed individual qubit readouts via a common feedline~\cite{Jerger12,Chen12}.  Finally, flux-bias lines allow individual tuning of qubit transition frequencies with $1~\ns$ resolution~\cite{DiCarlo09}.

The interaction step of the parity meter involves two controlled-phase (\cphase) gates between $\QAn$ and the data qubits. We compile these gates using a toolbox of resonant qubit--bus interactions proposed in Ref.~\onlinecite{Haack10b} and first realized with phase qubits~\cite{Mariatoni11}. A map of coherent qubit--bus interactions in the one- and two-excitation manifolds is obtained by varying the duration and amplitude of a flux pulse on $\QDi$ starting from $\ket{\eDi,0}$ and $\ket{\eDi,1}$, respectively [Fig.~\ref{Fig1}(b)]. We use $\ket{g_i}$, $\ket{e_i}$, and $\ket{f_i}$ to denote the ground, first, and second excited states of transmon $i$, respectively, and $\ket{n}$ to refer to the $n$-photon state of the bus. A half-period of oscillation at the $\ket{\eDi,0} \leftrightarrow \ket{\gDi,1}$ resonance [Fig.~1(c)] implements an \iswap\ gate~\cite{Johansson06} between $\QDi$ and $B$. A full period at the $\ket{\fDi,0} \leftrightarrow \ket{\eDi,1}$ resonance in the two-excitation manifold [Fig.~1(d)] implements a \cphase\ gate~\cite{Strauch03}.  We implement \cphase\ gates between $\QAn$ and $D_i$ using three qubit--bus primitives: \iswap$_{A,B}$, \cphase$_{B,Di}$ and \iswap$_{A,B}$. Note that the $n$ \cphase\ gates in the interaction step of an $n$-qubit ancilla-based parity measurement can be realized with only $n + 2$ qubit-bus primitives, since back-to-back $\QAn$-$B$ swaps can be compiled away ($n=2$ here).
 
\begin{figure}
\includegraphics[width=\columnwidth]{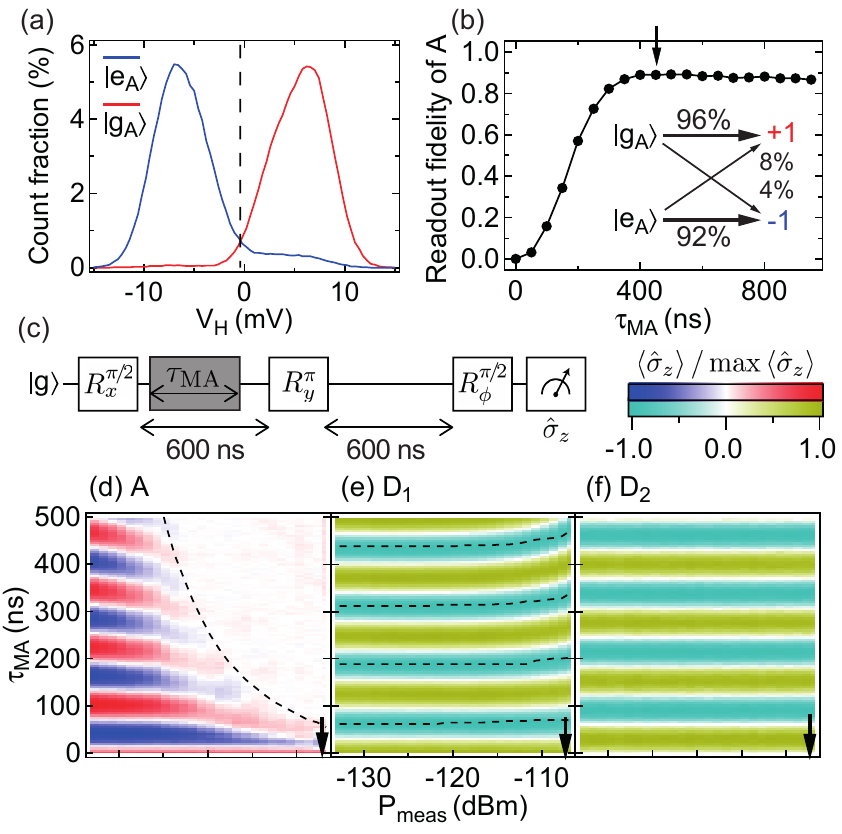}
\caption{
(a) Histograms of $\VHA$ ($\tauMA = 450\,\unit{ns}$) for computational states of $\QAn$. The dashed line is the fidelity maximizing digitizing threshold. (b) Single-shot ancilla readout fidelity $\Fa$ as a function of the measurement pulse duration $\tauMA$. Maximal $\Fa = 89\%$ is attained at $\tauMA = 450\,\unit{ns}$ (arrow). Inset: corresponding calibrated readout error model. (c) Gate sequence used to study qubit dephasing induced by $\QAn$ measurement. A readout pulse of duration $\tauMA$ and power $P_{\mathrm{meas}}$ is embedded in a fixed-length echo sequence performed on qubit $\QAn$ (d), $\QDi$ (e), and $\QDii$ (f). The azimuthal angle $\phi$ of the final $\pi/2$ rotation is swept from $0$ to $8\pi$ jointly with $\tauMA$ to facilitate discerning deterministic phase shifts and dephasing. The plots show averaged $\hat{Z}$ measurements normalized to compensate for the fixed loss of contrast due to intrinsic decoherence. In panel (d), the dashed line is a theoretical prediction for $98\%$ loss in phase contrast. Dashed lines in (e) are equal-phase contours accounting for AC Stark shift on $\QDi$. All theory curves were calculated using measured parameters~\cite{SM}. No effect on $\QDii$ is observed [panel (f)]. Arrows indicate the power used for $\QAn$ readout in Figs.~3 and 4. The incident power corresponding to 1-photon average population in the $\QAn$ readout resonator is $-133~\dBm$.
}
\label{Fig2}
\end{figure}
The ideal projective ancilla measurement comprising the second step of the parity meter is high fidelity, fast relative to intrinsic qubit decoherence, and does not impose any additional back-action on data qubits. We probe the ancilla-state-dependent transmission of a dedicated, dispersively-coupled resonator~\cite{SM} with a microwave pulse applied to the feedline near the resonator's fundamental (7.366~GHz). Following increasingly standard practice in circuit QED~\cite{Johnson12,*Riste12}, we use a Josephson parametric amplifier (JPA) at the front end of the amplification chain to boost the readout fidelity and reduce the pulse duration~\footnote{We were unable to tune the JPA resonance frequency high enough to align with $\QAn$'s resonator, which limited small-signal gain to $4.2~\dB$.}. Histograms of the integrated homodyne voltage $\VHA$ with ancilla prepared in $\ket{\eAn}$ and $\ket{\gAn}$ reveal an optimal single-shot fidelity of $89\%$ at measurement pulse duration $\tauMA = 450~\ns$ while probing with $\sim400$ intra-resonator steady-state photons [Fig.~2(a) and (b)]. Crucially, the ancilla measurement does not induce any significant dephasing on data qubits, despite the high level of measurement power used. To show this, we embed ancilla readout pulses in the first half of standard echo experiments on $\QAn$, $\QDi$ and $\QDii$ [Figs.~2(c)-(f)]. For $\QAn$, the expected coherence loss due to measurement is observed at all readout powers. For $\QDi$ ($\QDii$), only $2.4\,\pm\,0.6$\% ($2.1\,\pm\,1 \%$) of contrast is lost at the chosen measurement strength. On $\QDi$, whose readout resonator is closest in frequency to that of $\QAn$, we observe power-dependent qubit detuning consistent with the AC Stark shift~\cite{Schuster05}. We correct the induced deterministic phase either with a 5~ns detuning flux pulse, or in post-processing. To completely test the QND character of ancilla measurement, we perform quantum process tomography (QPT) on the data qubits undergoing $326~\ns$ of idling, with and without an applied 300~ns ancilla readout pulse~\cite{SM}. The 0.97 process fidelity between these two processes, after correcting for the phase accrued by $\QDi$ with measurement on, confirms the low level of back-action.

\begin{figure}
\includegraphics[width=\columnwidth]{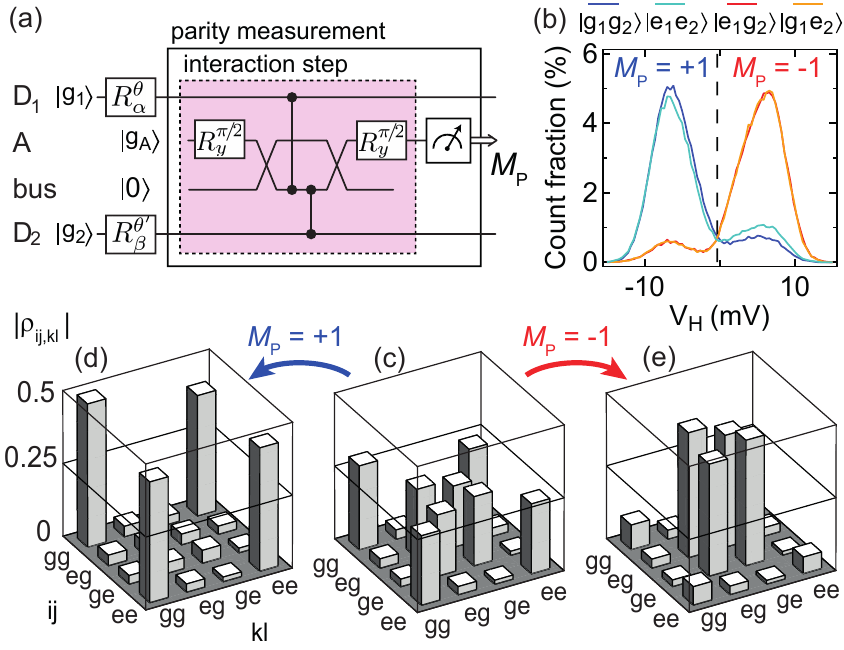}
\caption{(a) Gate sequence realizing an indirect parity measurement of $\QDi$ and $\QDii$ using coherent interactions with $\QAn$ and a subsequent projective measurement of $\QAn$. (b) Histograms of the integrated voltage $\VHA$ ($\tauMA = 450\,\unit{ns}$) for computational basis state inputs prepared using initial rotations $R_\alpha^\theta, R_\beta^{\theta'} \in \{I, R_x^\pi\}$. Digitizing threshold indicated by the dashed line produces the maximum parity fidelity $\mathcal{F}_{\txt{P}} = 69\%$. \hbox{(c)-(e)} Manhattan-style plots of data-qubit density matrices after the parity measurement for the separable superposition input state prepared with $R_\alpha^\theta = R_\beta^{\theta'} = R_y^{\pi/2}$. In the unconditioned tomogram (c), the density matrix elements indicative of coherences between the parity subspaces have been suppressed by 90\% in magnitude. The spurious residual probability amplitudes can be attributed almost entirely to the infidelity of the two \iswap\ operations. The conditioned tomograms (d) and (e) demonstrate probabilistic entanglement by measurement, reaching 87\% (81\%) fidelity to the even (odd) Bell state and 0.74 (0.63) concurrence for the even (odd) outcome. The higher fidelity of the even projection is in accordance with the error model of ancilla readout, where the dominating error mechanism is relaxation of $\QAn$ during readout. For (c)-(e), the timing of tomographic pre-rotations and measurements corresponds to the parallel variation with $\tMD = 0$ as illustrated in Fig.~\ref{Fig4}(b).
}
\label{Fig3}
\end{figure}

We now combine the interaction and measurement steps described into the full parity measurement protocol shown in Fig.~3(a). We first quantify the parity measurement fidelity by analyzing the correlation between measurement results $\Mp=\pm1$ for data-qubit input states of definite parity, namely the four computational states. The optimal digitizing threshold maximizes the parity readout fidelity at $\mathcal{F}_\txt{P} = 69\%$ [Fig.~3(b)]. To test the meter's ability to preserve (suppress) coherence within (across) parity subspaces, we apply parity measurement to the maximal but separable superposition state $\frac{1}{2} \left(\ket{\gDi \gDii} + \ket{\gDi \eDii} + \ket{\eDi \gDii} + \ket{\eDi \eDii} \right)$ created using two $\pi/2$ pulses. State tomography of the data qubits at the end of the interaction step [parallel timing, Fig.~3(c)] shows that the average absolute coherence between states of different parity [$\left(|\rho_{ge,gg}| + |\rho_{ge,ee}| + |\rho_{eg,gg}| + |\rho_{eg,ee}|\right)/4$, where $\rho_{ij,kl} = \bra{i_1 j_2} \rho \ket{k_1 l_2}$] is suppressed by $90 \pm 1\%$, while the average intra-parity absolute coherence $\left(|\rho_{ee,gg}| + |\rho_{eg,ge}|\right)/2$ decreases only $10 \pm 1\%$. Similarly, state tomography at the end of measurement step (serial timing) shows a total intra-parity coherence loss of $32 \pm 1\%$, consistent with intrinsic qubit decoherence during $\tauMA$. For parallel timing, conditioning on  $\Mp=+1(-1)$ unveils highly-entangled states with concurrence 0.74 (0.63) and Bell-state fidelity 87\% (81\%). The corresponding density matrices are shown in Figs.~3(d) and (e), respectively. For serial timing, these values reduce to 0.46 (0.38) and 73\% (67\%), respectively.

QPT of the data qubits with and without conditioning on the $\Mp$ outcome~\cite{SM} provides the most complete characterization of the parity measurement. For parallel timing, the fidelities to the corresponding ideal process are 0.91, 0.84, and 0.79 for no $\Mp$ conditioning, conditioning on $\Mp=+1$, and conditioning on $\Mp=-1$, respectively. For serial timing, the respective  process fidelities are  0.77, 0.70, and 0.65. From the process tomograms, we determined that the dominant error in the coherent interaction step was the 89\% population transfer efficiency of the \iswap\ gate between ancilla and bus~\footnote{The population transfer fidelity was limited by a spurious resonance 600~MHz above the bus resonator and strongly coupled to $\QAn$.}.

\begin{figure}
\includegraphics[width=\columnwidth]{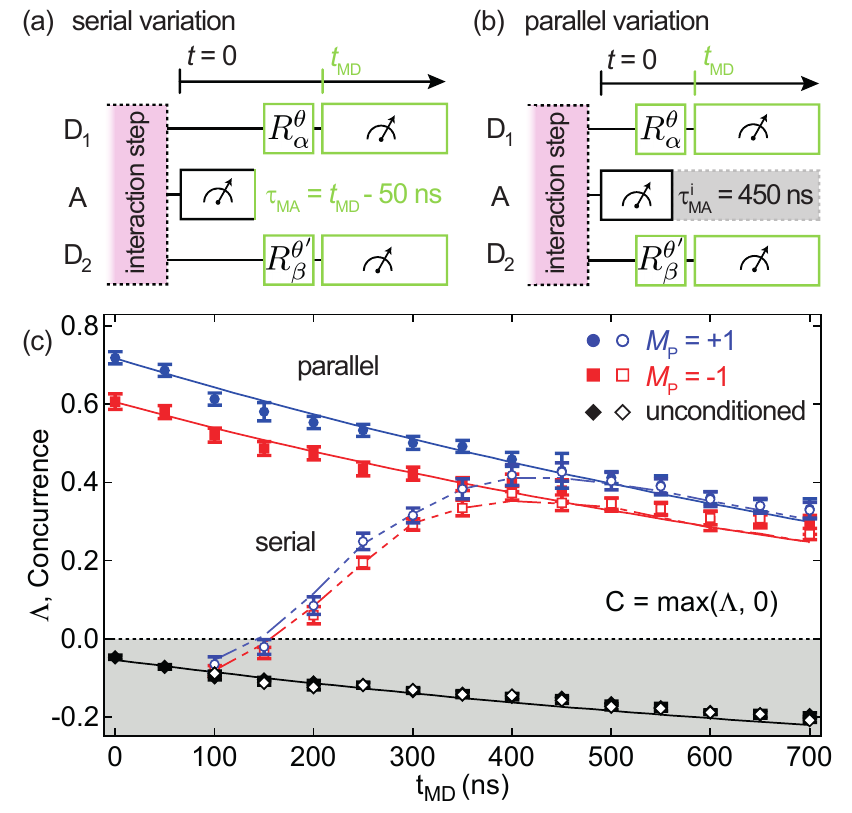}
\caption{Genesis and decay of two-qubit entanglement as a function of the time $\tMD$ between the end of the interaction step and the beginning of the readout pulses on the data qubits. (a) Timing diagram for the serial variation, in which the ancilla measurement is completed strictly before operations on data qubits continue. (b) Timing diagram for the parallel variation, in which the duration $\tauMA$ of the ancilla readout pulse is constant $1\,\unit{\mu s}$. The homodyne voltage is integrated only during the first $\tauMA^i = 450\,\unit{ns}$. The tomographic pre-rotations and the readout pulses for the data qubits can overlap with the ancilla readout. For $t_\txt{MD} = 0$, the pre-rotations are done simultaneously with the ancilla $R_y^{\pi/2}$ gate at the end of the interaction step. In (a)-(b), elements drawn in green move together in time. (c) Wootters' $\Lambda$~\cite{Wootters98} and concurrence $C = \max(\Lambda, 0)$ as a function of $\tMD$ for the parallel and serial timings (filled and open markers, respectively) extracted from the two-qubit density matrix that is unconditioned (black diamonds), conditioned on the even outcome ($\MP = +1$, blue circles), and conditioned on the odd outcome ($\MP = -1$, red squares) of the ancilla measurement. Solid curves are based on a model that includes single-qubit relaxation and dephasing processes, and uses the experimental density matrix at $t = 0$ as the initial value. For the dashed curves, mixing of the parity subspaces according to calibrated readout errors is explicitly included.
}
\label{Fig4}
\end{figure}

Finally, we study the competition between parity readout fidelity and intrinsic qubit decoherence in the entanglement-by-measurement protocol. We vary the idling time $\tMD$ between the end of the interaction step and the beginning of the data qubit readout pulse for both serial and parallel timings [Figs.~4(a) and 4(b), respectively]. For serial timing [Fig.~4(c), open markers], we use $\tauMA = \tMD - 50~\unit{ns}$, resulting in a steep initial increase in concurrence owing to rapidly improving ancilla readout fidelity followed by a decay due to intrinsic data qubit decoherence. To quantify the evolution from a product to an entangled state of data qubits, we consider Wootters' $\Lambda$~\cite{Wootters98} used to define concurrence $C(\rho) \equiv \max\{\Lambda(\rho), 0\}$. Even though the initial maximal superposition state lies at the boundary between separable and entangled two-qubit states, decoherence in the data qubits pulls the state away from the boundary, as manifested by the negative $\Lambda$ observed without conditioning on the ancilla measurement [Fig.~4(c), black markers]. This pull imposes a minimum threshold in the ancilla readout fidelity to generate entanglement by conditioning on the measurement outcome. Entanglement is established after $\tauMA \approx 100~\unit{ns}$. This time is quantitatively matched by a model including the calibrated ancilla readout errors. For parallel timing [Fig.~4(c), filled markers], in which we use the optimal integration time $\tau_{MA}^i = 450~\unit{ns}$ for all $\tMD$, entanglement decreases monotonically and consistently with the intrinsic qubit decoherence. For $\tMD > 450~\ns$, the parallel and serial timings perform similarly, because the ancilla readout fidelity is nearly constant.

The above entanglement by measurement is a discretized version of the continuous-time scheme investigated theoretically in Ref.~\onlinecite{Williams08b}. The finite time to entanglement observed in serial timing is reminiscent of the entanglement genesis time required under continuous parity measurement. However, while the continuous scheme produces entanglement even starting from a maximally mixed state owing to the interplay of simultaneous Hamiltonian and measurement dynamics,  entanglement by a discrete, projective parity measurement necessitates an initial superposition state of the data qubits. Instead, performing two parity measurements with single-qubit rotations in between would realize a QND Bell-state measurement~\cite{Ionicioiu07}, producing entanglement for any input two-qubit state. This protocol could be conveniently implemented with this processor in parallel timing by employing the unused qubit as a second ancilla.

In conclusion, we have realized a two-qubit parity meter in 2D cQED using a two-step scheme involving interaction of the data qubits with an ancilla and subsequent ancilla projection. The interaction step, employing resonant interactions at the raw speed set by qubit-bus coupling, can be efficiently compiled into $n+2$ primitives for $n$-qubit parity measurement. Detailed characterization of the ancilla readout performed via a dedicated dispersively-coupled resonator demonstrates minimal measurement-induced dephasing of data qubits (97\% of single-qubit coherence retained), low measurement cross-talk (2\% during simultaneous three-qubit readout)~\cite{SM} and high single-shot fidelity (89\%).  Applying the parity measurement on an unentangled superposition state of the two data qubits generates entanglement for both measurement outcomes, in both serial and parallel timings. In the former, we observe entanglement genesis after a 100~ns ancilla measurement. As a possible follow-up experiment, coupling a fifth qubit to the bus would allow implementing the four-qubit parity measurements necessary for quantum error correction using surface codes. We anticipate that the enhanced 2D+ connectivity offered by recent fabrication developments~\cite{Steffen13, Chen13} will also allow implementing larger fragments of error-correcting lattices using this architecture.

\begin{acknowledgments}
\emph{Contributions and acknowledgements.} O.P.S. designed and fabricated the processor based on earlier devices by J.P.G. and M.M., did the measurements and data analysis, and wrote the manuscript with L.D.C. J.C. and J.P.G. devised the two-qubit \cphase\ tuneup. G.d.L. designed the mechanical parts used in the low-temperature setup. L.D.C. initiated and supervised the project. We thank W.~Kindel and K.~W.~Lehnert for the JPA, D.~Thoen and T.~M.~Klapwijk for NbTiN thin films, D.~Rist\`e for assistance with measurements, and A.~N.~Jordan for discussions. We acknowledge funding from the Netherlands Organization for Scientific Research (NWO, VIDI scheme) and the EU FP7 projects SOLID and SCALEQIT.
\end{acknowledgments}

\bibliographystyle{prl}
\bibliography{../../../TeX/References/References_cQED} 

\end{document}


\title{Supplemental material for ``Entanglement genesis by ancilla-based parity measurement in 2D circuit QED"}
\author{O.-P.~Saira}
\author{J.P.~Groen}
\author{J.~Cramer}
\author{M.~Meretska}
\author{G.~de~Lange}
\author{L.~DiCarlo}
\affiliation{Kavli Institute of Nanoscience, Delft University of Technology, P.O. Box 5046,
2600 GA Delft, The Netherlands}
\date{\today}

\maketitle

\section{Quantum processor}

\emph{Fabrication.} The quantum processor was fabricated using a flow similar to Ref.~\cite{Groen13}. The substrate is a C-plane sapphire wafer (thickness 430~$\mu$m), on which the coplanar waveguide transmission lines and the ground-plane grid were defined by reactive-ion etching of a NbTiN film (thickness 80~nm). The transmons were defined using double-angle evaporation of aluminum (thicknesses 15~nm and 25~nm for the bottom and top layers, respectively) and in-situ oxidation to realize the tunnel junctions (0.8~mbar O$_2$ atmosphere for 480~s).

\emph{Characterization.} The measured characteristics of the quantum elements of the chip are presented in Tbl.~\ref{Tbl:SOM-params}. The Josephson energy $\EJ$ of each qubit, and hence also the transition frequency $f_{ge}$, was individually tunable. Maximum $f_{ge}$ denotes the transition frequency at the `sweet spot'. The values for charging energy $\EC$ and for $\EJ$ were obtained by fitting them to the measured $f_{ge}$ and $f_{ef}$ at one bias point using a numerically exact model for the transmon spectrum~\cite{Koch07}. During the execution of gate sequences, the qubit frequencies were tuned to the operation points indicated in the table. The fourth, unused qubit was tuned to $5.42~\unit{GHz}$. We used a pulsed flux-biasing scheme where the duration of the biasing pulse was typically $2~\unit{\mu s}$, except for the coherence time measurements, where the bias pulse started $2~\unit{\mu}s$ before the first rotation pulse. The coherence times of the bus were measured by swapping an excitation into and out of the bus using $\QDi$. The coupling strengths $g$ with the bus were determined from the observed vacuum Rabi oscillation periods, and agree with the avoided crossing observed in spectroscopic measurements. The cQED parameters describing a readout resonator coupled to a transmon and the feedline were extracted from standard spectroscopic measurements. The $1$-photon power refers to the incident power at the feedline producing 1-photon average intra-resonator population in steady state, estimated by measuring the AC Stark shift~\cite{Schuster05} induced on the qubit with small photon numbers ($\bar{n} \lesssim 10$) using a separately calibrated value for the dispersive shift $\chi$.

\begin{table}[!h]
\caption{Summary of the main device parameters.}
\label{Tbl:SOM-params}
\begin{tabular}{l | c c c | c}
  				    	& $\QAn$ 	& $\QDi$ 	& $\QDii$ 	& bus \\
\hline
max $f_{ge}$, GHz 		& 5.878 	& 6.812	& 6.530	&\\
max $\EJ/h$, GHz			& 15.3	& 23.1	& 19.9	&\\
$\EC/h$, GHz			& 0.31	& 0.27	& 0.29	&\\
\hline
operation point $f_{ge}$, GHz	& 5.878	& 6.812 	& 6.340	& 4.958\\
$T_1$, $\mu$s 			& 6.5		& 3.6		& 6.3		& 4.3\\
$T_2$, $\mu$s			& 1.3		& 3.6		& 2.1		& 8.1\\
$T_2^{\txt{echo}}$, $\mu$s	& 7.1		& 3.5		& 5.9		& -\\
\hline
$g/2\pi$ to bus, MHz		& 9.6		& 19.1	& 19.5	\\
\hline
bare $f_r$, GHz			& 7.364	& 7.421	& 7.469\\
$\chi/\pi$, MHz			& $-$0.7	& $-$4.4	& $-$1.6\\
$g/2\pi$, MHz			& 55		& 66		& 67\\	
$\kappa/2\pi$, MHz			& 1.1		& 1.3		& 1.7\\
1-photon power (dBm)		& $-$133	& $-$138	& $-$133
\end{tabular}
\end{table}

\section{Experimental setup}

\begin{figure*}
\includegraphics[width=\textwidth]{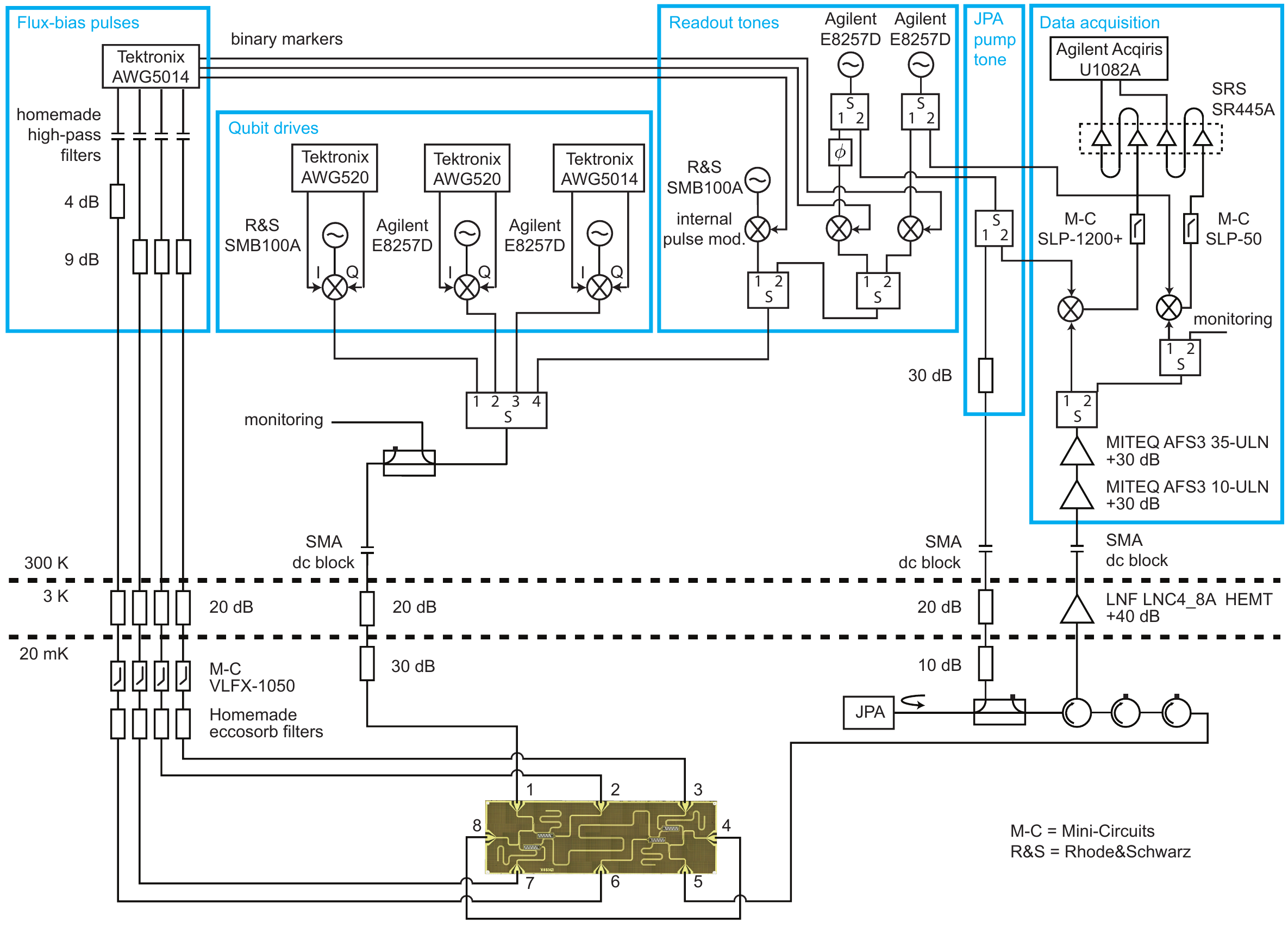}
\caption{Complete wiring schematic.}
\label{Fig:SOM-wiring}
\end{figure*}

\emph{Wiring.} The quantum processor was cooled to 22~mK using a dilution refrigerator. A complete schematic of the experimental wiring and hardware used for pulse generation and data acquisition is illustrated in Fig.~\ref{Fig:SOM-wiring}. The qubit rotations were quadrature-modulated pulses with a Gaussian envelope (total duration $t_g = 4\sigma = 24~\unit{ns}$), augmented with the `DRAG' scheme of Ref.~\onlinecite{Motzoi09} to reduce the leakage to $\ket{f}$ level.

\begin{figure}
\includegraphics[width=\columnwidth]{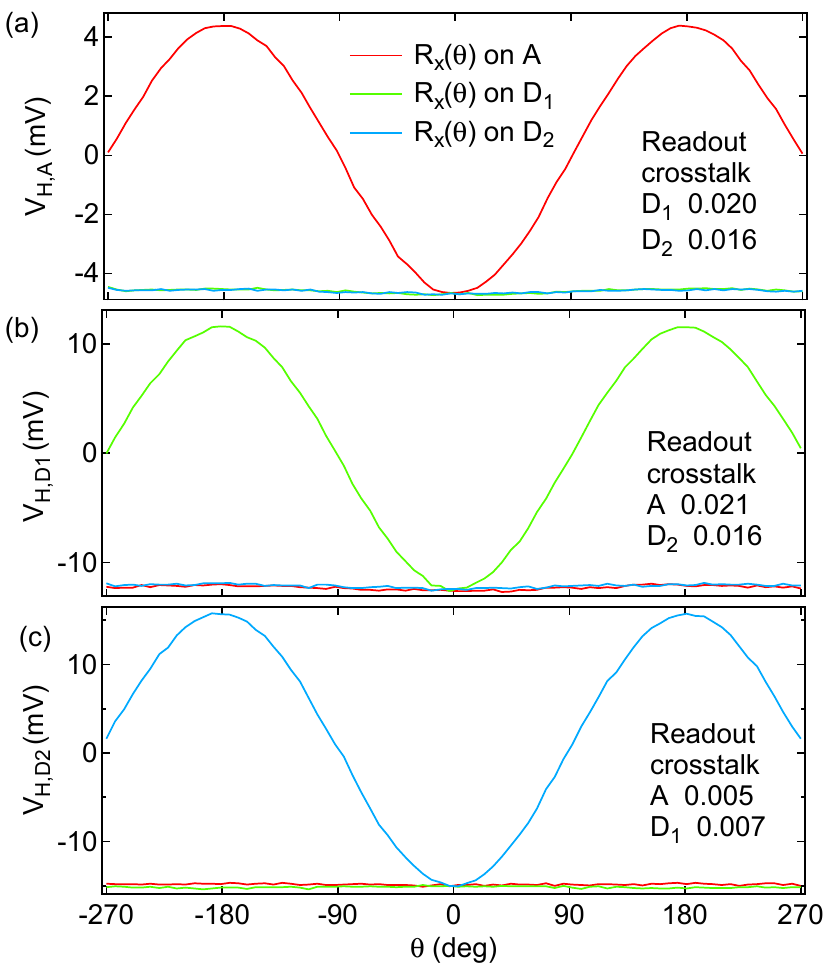}
\caption{Averaged integrated voltages $\left<V_{\mathrm{H},j} \right>$ from multiplexed three-qubit readout for different single-qubit state preparations. Starting from the thermal ground state, a rotation pulse $R_x(\theta)$ was applied to one of the qubits $\QAn$, $\QDi$, or $\QDii$ (see legend), after which simultaneous readout tones were generated for all three readout resonators. We characterize the visibility of a qubit in a particular readout by the amplitude of the observed Rabi oscillation signal. The quoted crosstalk of $i$ in the readout $\left<V_{\mathrm{H},j} \right>$ is the ratio of the fitted amplitude resulting from $i$ rotation to that resulting from $j$.}
\label{Fig:SOM-arches}
\end{figure}

\emph{Multiplexed readout.} The three readout resonators used could be addressed independently as their frequency spacing is much larger than any of the linewidths, and also much larger than the inverse of the readout pulse duration. Furthermore, we expect negligible state-dependent frequency shifts of resonators from qubits not coupled to them. Nonetheless, the experimental homodyne voltages do show few-percent cross-talk, which could also arise from non-linearity in the readout chain. Figure~\ref{Fig:SOM-arches} shows the dependence of the three integrated homodyne voltages on the state of each qubit using simultaneous, frequency-multiplexed measurement.

\section{Extended results}

\begin{figure}[h]
\includegraphics[width=\columnwidth]{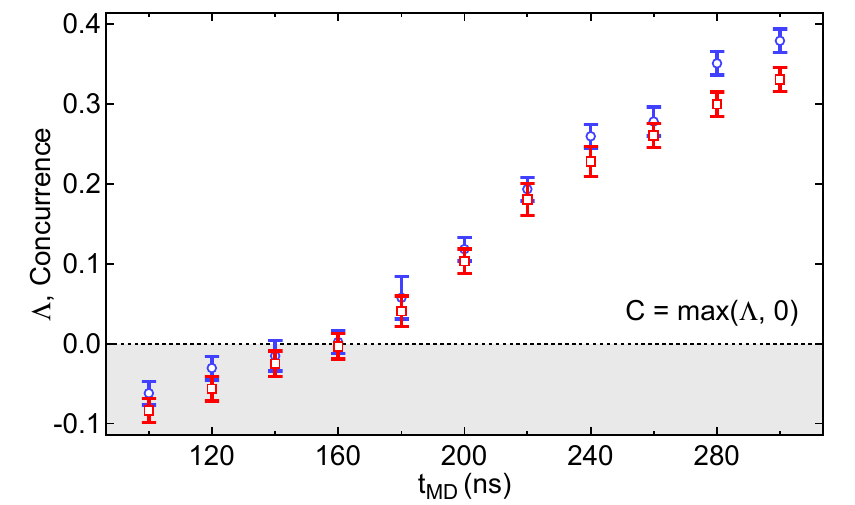}
\caption{A plot similar to Fig.~4(c) of the main text, where we show Wootter's $\Lambda$ and concurrence for the two-data-qubit state after parity measurement conditioned on the even (blue) and odd (red) outcomes in the `serial' timing variation. The range of $t_\txt{MD}$ values was chosen as to illustrate more clearly the entanglement genesis. For this dataset, we used a slightly higher JPA pump power and omitted the first 80~ns of the homodyne signal when calculating the integrated voltage. For the short readout pulses considered here, this readout configuration resulted in slightly better $\QAn$ readout fidelities and correspondingly higher concurrences. However, using this configuration with longer pulses resulted in double-peaked histograms, possibly indicating that the JPA was bifurcating.
}
\label{Fig:SOM-lambda}
\end{figure}

In this section, we present datasets supporting claims made in the main text. Fig.~\ref{Fig:SOM-lambda} shows a second dataset highlighting the entanglement genesis observed in the entanglement-by-measurement experiment in serial timing.

\begin{figure*}[h]
\includegraphics{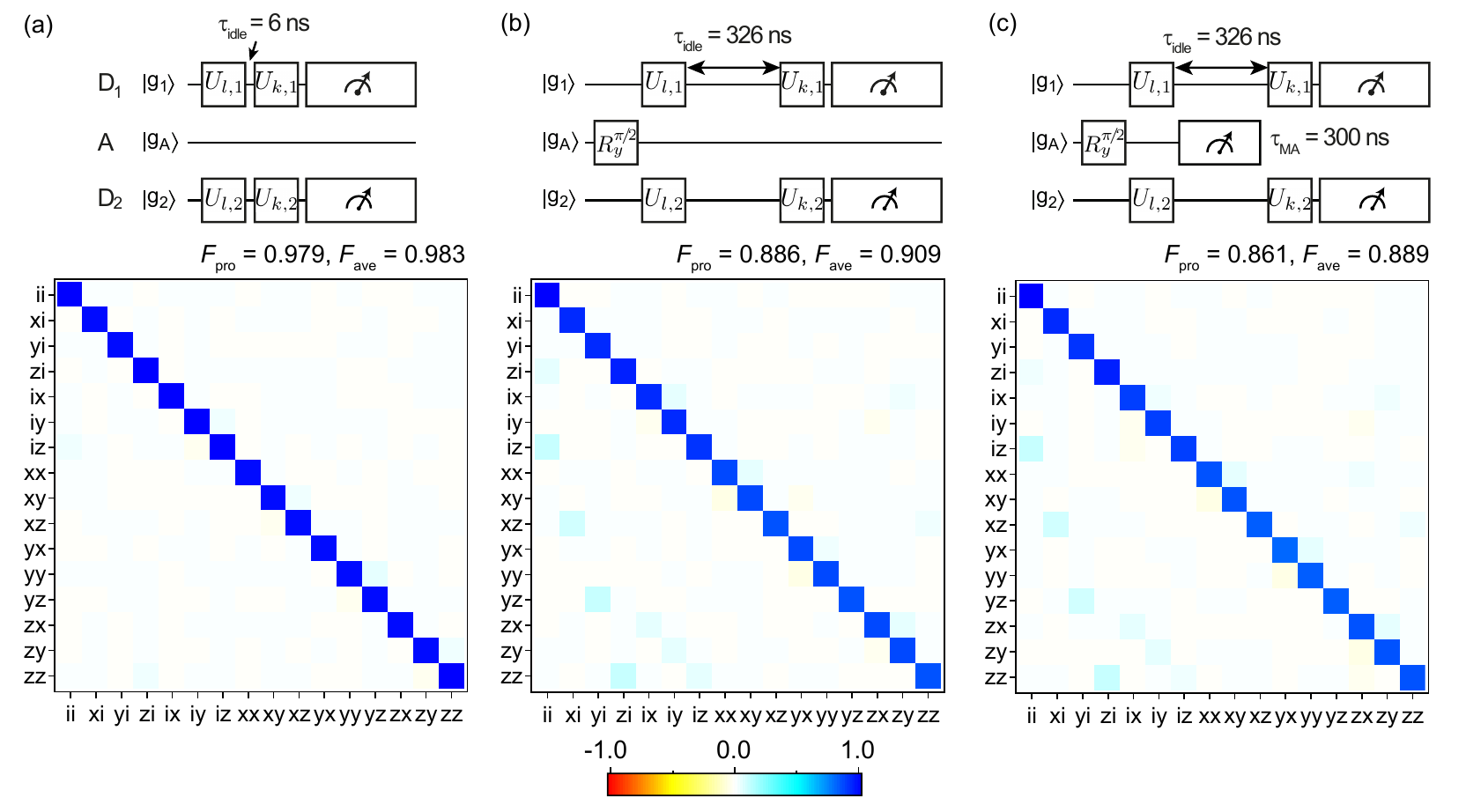}
\caption{Process tomograms in the $\mathcal{R}$ matrix representation in the data qubit subspace for three scenarios in which the target operator is the identity. (a) Logical identity. Here, the state preparation and tomographic pre-rotation pulses were only separated by the $6~\ns$ buffer used in all pulse synthesis. (b) $326~\ns$ idling gate. (c) 326~ns idling gate with a $300~\ns$ ancilla measurement tone applied during the waiting period. In (b) and (c), we have corrected in post-processing the determinstic single-qubit phases due to small detunings of the qubit drive frequencies and the AC Stark shift [affecting only (c)]. The similarity of tomograms (b) and (c), characterized by a mutual process fidelity $\Fpro = 0.966$, indicates that a projective ancilla measurement perturbs only weakly the state of the data qubits.
}
\label{Fig:SOM-qpt-id}
\end{figure*}

\begin{figure*}[h]
\includegraphics{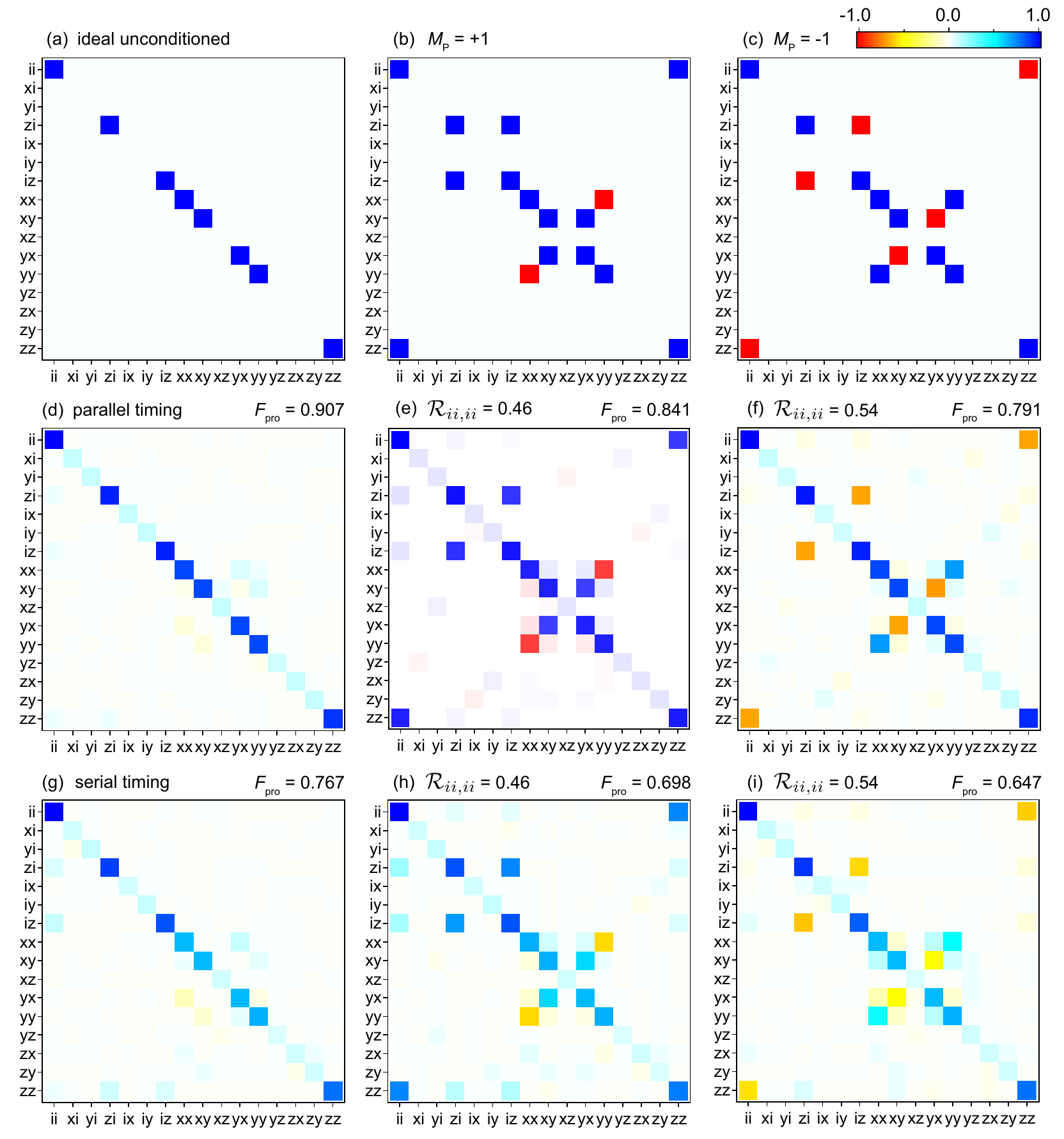}
\caption{Process tomograms for the parity measurement in the data qubit subspace. \emph{Row-wise:} (a)-(c) Ideal result. (d)-(f) Experiment with `parallel' timing. (g)-(i) Experiment with `serial' timing. \emph{Column-wise:} (a), (d), (g) Evolution without conditioning on $\MP$. The unconditioned evolution is trace-preserving but non-unitary as coherence across the parity subspaces is suppressed.  (b), (e), (h) Evolution conditioned on the even outcome $\MP = +1$. (e), (f), (i) Evolution conditioned on the odd outcome $\MP = -1$. Evolution that is conditioned on $\MP = P$, $P \in \left\{-1,1\right\}$, can be understood as a non-trace preserving map $\rho \mapsto \rho'$, where $\tr(\rho')$ is the probability to obtain $\MP = P$ for input $\rho$. For visualization, the $\mathcal{R}$ matrices have been normalized by the $(ii, ii)$ matrix element [equal to $\tr(\chi)$ in $\chi$-matrix representation] that tells the probability of this $\MP$ outcome for the completely mixed two-qubit density matrix $\rho = I / 4$. The unscaled $R_{ii,ii}$ values are indicated above the plots.
}
\label{Fig:SOM-qpt-parity}
\end{figure*}

Figure \ref{Fig:SOM-qpt-id} shows process tomograms for three idling variations. QPT for 6~ns of idling in panel (a) was used to benchmark the accuracy of the QPT protocol, and to extract single-qubit rotation errors in a self-consistent manner. QPTs for 326~ns of idling without and with ancilla readout in panels (b) and (c), respectively, show the low level of back-action resulting from the projective ancilla readout. The set of tomograms in Fig.~\ref{Fig:SOM-qpt-parity} constitutes a full characterization of the parity measurement as a quantum circuit element. The tomograms allow the calculation of the process fidelity values quoted in the main text, and also help in identifying the dominant error processes.

\section{Hamiltonian model and error processes}

The processor has nine quantum elements: four frequency-tunable transmon qubits, four readout resonators each coupled to a qubit and to the shared feedline, and one bus resonator coupling to all qubits. To understand the ideal coherent operation of the parity measurement and the dominant non-idealities, it is sufficient to consider the following subsystems in isolation.

\emph{Qubit \& bus resonator.} The Hamiltonian of a subsystem consisting of qubit $k$ and the bus is
\begin{equation}
H = \hbar \omega_B a^\dag a + \sum_{j \ge 0} \hbar \omega_j^{(k)} \ket{j_k} \bra{j_k} + H_I,
\end{equation}
where $\omega_B$ is the fundamental bus resonance frequency, $a^\dag$ ($a$) is the photon creation (annihilation) operator for the bus, $\hbar \omega_j^{(k)}$ is the energy of the $j$th transmon level, and $H_I$ is the coupling term. In the transmon regime $\EC \ll \EJ$, the coupling assumes the form
\begin{equation}
H_I = \hbar g_0^k \sum_{j \ge 0} \sqrt{j + 1} \left( \ket{j+1_k} \bra{j_k} a + \mathrm{h.c.} \right),
\end{equation}
where $2 g_0^{(k)}$ is the vacuum Rabi splitting between the qubit and the bus. To describe the ideal implementation of the \cphase\ gate, we consider two cases of resonant coupling: At detuning $\omega_1 - \omega_0 = \omega_B + \Delta \omega$, the interaction picture Hamiltonian reads
\begin{equation}
H^1_\txt{int} = e^{i  \Delta \omega\,t}\hbar g_0^{(k)} \ket{e_k, 0} \bra{g_k, 1} +\mathrm{h.c.},
\end{equation}
where we have dropped the other, fast-oscillating terms. Evolution at resonance $\Delta \omega = 0$ for time $t_\txt{swap} = \pi / (2 g_0^{(k)})$ implements a coherent swap between $\ket{e_k, 0}$ and $\ket{g_k, 1}$. The acquired single-qubit phases will be accounted for later, so that only the population transfer is relevant. Experimentally, the challenge is to realize a non-adiabatic tuning to $\Delta \omega = 0$ starting from the non-interacting operation point of the qubit, where $\Delta \omega/ (2 \pi) \sim 1\ldots2\ \unit{GHz}$, see Tbl.~\ref{Tbl:SOM-params}. Inaccuracy in the timing or amplitude of the detuning pulse will limit the population transfer. However, the transfer probability is only second-order sensitive to both $\Delta \omega$ and $t - t_\txt{swap}$.

Similarly, at detuning $\omega_2 - \omega_1 = \omega_B +  \Delta \omega$, the Hamiltonian in the interaction frame reads
\begin{equation}
H^2_\txt{int} = e^{i  \Delta \omega\,t} \sqrt{2} \hbar g_0^{(k)} \ket{e_k, 1} \bra{f_k, 0} + \mathrm{h.c.},
\end{equation}
again keeping only the slowest-oscillating terms. Evolution at $\Delta \omega = 0$ for time $t_\txt{cp} = \pi /(\sqrt{2} g_0^{(k)})$ ideally leaves behind no population in the bus, but the $\ket{e_k, 1}$ state acquires an additional phase shift of $\pi$ in the interaction frame. Hence, this interaction implements a \cphase\ gate~\cite{Haack10b} between the bus and the qubit, modulo single-qubit phases. For an experimental \cphase\ realization, one needs to consider errors both in population transfer (second-order sensitive to $\Delta \omega$ and $t - t_\txt{cp}$) and the conditional phase (first-order sensitive). Here, population transfer errors leave the qubit potentially in the $\ket{f_k}$ level, constituting leakage from the computational subspace.

\emph{Qubit \& readout resonator.}
The subsystem consisting of a transmon qubit and its readout resonator is described in the dispersive limit $\left| \omega_{ge} - \omega_r \right| \gg g$ by the dispersive Jaynes-Cummings hamiltonian~\cite{Koch07}
\begin{equation}
H = \hbar \omega_r' a^\dag a - \frac{\hbar}{2} \omega_q' \sigma_z - \hbar \chi \sigma_z a^\dag a,
\end{equation}
where $\omega_r'$ and $\omega_q'$ are renormalized resonator and qubit frequencies, respectively, and $\chi$ is the dispersive shift. Measurement photons in the resonator have two effects on the qubit state, namely a shift in the qubit frequency and increased dephasing. In the dispersive limit, the evolution of the qubit density matrix under a measurement pulse is described by the model presented in Ref.~\cite{Gambetta06}. Disregarding intrinsic relaxation and dephasing of the qubit, the off-diagonal element $\rho_{eg}$ decays as
\begin{equation}
\rho_{eg}(t) = \rho_{eg}(0) \exp \left( -i 2 \chi \int_{0}^{t} \alpha_+(t') \alpha_-^*(t') dt' \right),\label{Eq:MIDeg}
\end{equation}
where $\alpha_\mp(t)$ describes the average resonator field corresponding to the ground (excited) state of the qubit, and can be solved from the  differential equation
\begin{align}
\dot{\alpha}_+(t)& = -i \epsilon_{\txt{rf}}(t) - i(\Delta_r + \chi - i \kappa/2) \alpha_+(t)\nonumber\\
\dot{\alpha}_-(t) &= -i \epsilon_{\txt{rf}}(t) - i(\Delta_r - \chi - i \kappa/2) \alpha_-(t),
\end{align}
where $\epsilon_{\txt{rf}}(t)$ is the amplitude of an external drive at $\omega_\txt{rf}$, and $\Delta_r = \omega_{\txt{rf}} - \omega'_r$. To convert the incident power $P_\txt{rf}$ into drive amplitude $\epsilon_\txt{rf}$, we use the relation
\begin{equation}
\frac{P_\txt{rf}}{P_\txt{1-ph}} = \frac{\epsilon_\txt{rf}^2}{\kappa^2/4},\\\label{Eq:MID1ph}
\end{equation}
where the power $P_\txt{1-ph}$ is calibrated using a continuous resonant measurement tone as discussed earlier. To model the measurement-induced dephasing and phase shifts observed in the experiment [Figs. 2(d)--(f)], we use Eqs.~(\ref{Eq:MIDeg})-(\ref{Eq:MID1ph}) to simulate the effect of a square-envelope measurement pulse at $\omega_{\txt{rf}} = \omega'_{r,A} + \chi$ on $\rho_{eg}$ including the decay of the resonator back to the vacuum state after the pulse.

\emph{Intrinsic decoherence.} Coupling of the computational subspace to the environment leads to decoherence. In the limit of a large number of weakly coupled environmental modes, the time evolution in the computational subspace is described by a master equation in Lindblad form. We will apply this dissipative equation to model the decay of data qubit coherence after the interaction step. Assuming that the interactions between the data qubits and other quantum elements in the processor are negligible at the operation point, the master equation reads~\cite{Gambetta06}
\begin{equation}
\dot{\rho} = \sum_{k = 1}^2 \gamma_{eg,k} \mathcal{D}[\sigma^-_k] \rho + \sum_{k = 1}^2 \gamma_{\phi,k} \mathcal{D}[\sigma_{z,k}] \rho,\label{Eq:Lblad}
\end{equation}
where the dissipation superoperator $\mathcal{D}[A] \rho = (2 A \rho A^\dag - A^\dag A \rho - \rho A A^\dag)/2$, and $\gamma_{eg,k}$ and $\gamma_{\phi,k}$ are the relaxation and pure dephasing rates, respectively, for qubit $k$.
Dissipative losses occurring during the interaction step are captured in the state and process tomograms taken with zero delay ($t_\txt{MD} = 0$), and we do not model them explicitly. Instead, we use the experimental density matrices as the initial conditions at $t = 0$, and study the decay of coherence according to Eq.~(\ref{Eq:Lblad}). Note that the above dissipative model for qubit dephasing does not fully describe the experimental conditions, since it does not capture refocusable phase errors. Nevertheless, the model is appropriate for studying free decay, which is the case here.

\section{State and process tomography}

In this section, we detail the protocol we used to perform state and process tomography in the data qubit subspace $\{\ket{g_1 g_2}, \ket{g_1 e_2}, \ket{e_1 g_2}, \ket{e_1 e_2} \}$. We first consider the case where ancilla measurement results are ignored. Our protocol closely follows the method presented in Ref.~\onlinecite{Chow12}.

\emph{Measurement model.} Using the multiplexed readout described earlier, each single-shot measurement yields two integrated homodyne voltages $\VHDi$ and $\VHDii$. In post-processing, we first subtract an offset voltage $\tilde{V}_i$ common to all measurements to obtain $\tilde{V}_{\mathrm{H},i} = V_{\mathrm{H},i} - \tilde{V}_i$. We construct three measurement operators $M_i$ whose expectation values $\langle M_i \rangle = \tr(M_i \rho)$ are experimentally determined as
\begin{align}
\langle M_1 \rangle &= \langle\langle \tVHDi \rangle\rangle\nonumber\\
\langle M_2 \rangle&= \langle\langle \tVHDii \rangle\rangle\nonumber\\
\langle M_3 \rangle &= \langle\langle \tVHDi \tVHDii \rangle\rangle,\nonumber
\end{align}
where double brackets denote averaging over repeated measurements. The most general form for the $M_i$ in dispersive cQED is~\cite{Filipp09}
\begin{equation}
M_i = \beta_{i0} + \beta_{i1} \sigma_z^1 + \beta_{i2} \sigma_z^2 + \beta_{i3} \sigma_z^1 \sigma_z^2,
\end{equation}
where the $\beta_{ij}$ are real coefficients. In the experiment, we calibrate the $\beta_{ij}$ by measuring the $\left< M_i \right>$ values for the four computational basis states.

\emph{State tomography.} To determine $\rho$, we precede the measurement step by tomographic pre-rotations $U_k$ chosen from the set $\mathcal{U} = \{I, R_x^\pi, R_x^{\pi/2}, R_x^{-\pi/2}, R_y^{\pi/2}, R_y^{-\pi/2}\}^{\otimes 2}$. In this manner, we obtain a total of 3 $\times$ 36 = 108 averaged measurements $\bar{m}_{ik}$ that are related to $\rho$ via
\begin{equation}
\bar{m}_{ik} = \tr\left({ U_k^\dagger M_i U_k \rho}\right).
\end{equation}
We work in the Pauli basis, representing $\rho$ as $\rho = \sum_{n} p_n P_n$, where $p_n = \tr(P_n \rho)/4$ and $P_n \in \{I, \sigma_x, \sigma_y, \sigma_z\}^{\otimes 2}$. We fix $\tr(\rho) = 1$, reducing the number of unknown $p_n$ to 15. We then obtain an overdetermined set of 108 linear equations of the form
\begin{equation}
\sum_n \tr\left({U_k^\dagger M_i U_k P_n}\right) p_n = \bar{m}_{ik}\label{Eq:mik},
\end{equation}
which we solve by weighted least-squares inversion. Each equation is weighted by the inverse variance of the single-shot measurements from which the average $\bar{m}_{ik}$ on the r.h.s. is calculated. 

\emph{Process tomography.} A quantum channel $\mathcal{E}$ is a linear trace-preserving map of density matrices. Representing the input and output density matrices in the Pauli basis, the unknown channel becomes a real-valued 16 $\times$ 16 matrix $\mathcal{R}$ known as the Pauli transfer matrix~\cite{Chow12}. To perform QPT, we augment state tomography protocol above by adding state preparation steps to the beginning. We take the state-preparation rotations from the set $\mathcal{U}$ defined above, so that selecting $U_l \in \mathcal{U}$ prepares an input state $\rho_l = U_l \ket{g_1g_2} \bra{g_1g_2} U_l^\dag$. The averaged measurement of $M_i$ with state-preparation $U_l$ and pre-rotation $U_k$ is related to the $\mathcal{R}$ matrix as
\begin{equation}
\bar{m}_{ikl} = \sum_{nm} \mathcal{R}_{nm} \tr (U_k^\dag M_i U_k P_n) \bra{g_1g_2} U_l^\dag P_m U_l \ket{g_1g_2}.\label{Eq:mikl}
\end{equation}
To solve this equation group, we first extract the matrices $\rho'_l = \mathcal{E}(\rho_l)$ by least-squares inversion as before. Then, treating the $\mathcal{R}_{nm}$ elements as 16 $\times 16$ = 256 unknowns, we obtain a group of $36 \times 16 = 576$ linear equations
\begin{equation}
\sum_{m} \mathcal{R}_{nm} \bra{g_1g_2} U_l^\dag P_m  U_l \ket{g_1g_2} = \tr(P_n \rho_l'),\label{Eq:Rnm}
\end{equation}
one for each choice of $n$ and $l$, which we solve by a final unweighted least-squares inversion.

\emph{Conditioning on the ancilla.} To fully characterize the parity measurement protocol, we need to consider density matrix evolution conditioned on the binary outcome of ancilla measurement. Formally, we can model this as a three-qubit process $\rho \otimes \ket{g_A} \bra{g_A} \mapsto \rho'_o \otimes \ket{g_A} \bra{g_A} + \rho'_e \otimes \ket{e_A} \bra{e_A}$, where $\tr(\rho'_{o(e)})$ gives the probability for the odd (even) measurement outcome, respectively, for a given input density matrix $\rho$.

To extract the conditioned density matrices $\rho'_{o(e)}$ for such a process, we extended the state tomography protocol described above as follows: in addition to the data qubit readouts $\VHDi$, $\VHDii$, we also record the ancilla readout result $\VHA$. Using the fidelity-optimizing threshold for $\VHA$, we obtained the readout calibration coefficients $\beta_{ij}$ for $\MA = +1$ and $\MA = -1$. To obtain $\rho'_{o(e)}$ for a fixed input $\rho$, the $\bar{m}_{ik}$ on r.h.s. of Eq.~(\ref{Eq:mik}) are given by the mean of the $M_i$ shots for which the corresponding $\MA = \pm 1$. The density matrix obtained from the inversion step is multiplied by $P(\MA = \pm1)$, \ie, the total fraction of even (odd) measurement outcomes for this particular input $\rho$. To construct the conditioned $\mathcal{R}$ matrices, we use conditional density matrices obtained in the above manner on the r.h.s. of Eq.~(\ref{Eq:Rnm}).

\emph{Rotation errors.} Quantum state and process tomography methods are vulnerable to systematic errors in state preparation and measurement. Here, we consider a particular class of errors in the set of rotations $\mathcal{U}$ that we can calibrate and correct for. Because all dedicated resonators couple to the same feedline, a microwave drive addressing the $\ket{g}$--$\ket{e}$ transition of a particular qubit also acts as an off-resonant Rabi drive on the other qubits, effectively realizing unwanted $z$-rotations. In two-qubit state tomography, we allow each rotation to induce a phase shift on the qubit ideally left unaffected by the pulse. Formally, we add the terms $\alpha^{(1)}_k \sigma^{(1)}_z  + \alpha^{(2)}_k \sigma^{(2)}_z$ to the control Hamiltonian generating each rotation $U_k \in \mathcal{U}$. By simulating the effect of such a faulty rotation set on the QPT protocol, we find the set of $\alpha$ values that best reproduces (in the least-squares sense) the deviations from identity observed in the experimental QPT for 6~ns idling. To correct these errors in subsequent tomograms, we use these rotations when constructing the model equations for state and process tomography according to Eqs.~(\ref{Eq:mik}) and (\ref{Eq:mikl}).

\emph{Ensuring physicality.} Physical density matrices are hermitian, positive semidefinite, and have unity trace. The same conditions characterize a physical process matrix in $\chi$-matrix representation~\cite{Nielsen00} except that $\tr(\chi) = d$, where $d$ is the dimensionality of the Hilbert space ($d = 4$ here). The state and process tomography procedures described above can produce $\rho$ and $\chi$ matrices with negative eigenvalues due to systematic and statistical errors. Therefore, as a final step in the processing of tomography data, we use numerical optimization to find the hermitian, positive semidefinite matrix that is closest in least-squares sense to the `raw' output of the tomographic inversion and has the same trace. In more detail, for a given $A^\txt{raw}$ we consider the optimization problem
\begin{align}
\min & \sum_{ij} |A_{ij} - A_{ij}^\txt{raw}|^2\nonumber\\
\mathrm{s.t.} & \ A^\dag = A, A \ge 0, \tr(A) = \tr(A^\txt{raw}).
\end{align}
Following Ref.~\onlinecite{Chow12}, we parametrize $A$ as the sum of $\tr(A^\txt{raw}) I/d$ and a linear combination traceless hermitian basis matrices, transform the quadratic objective function into a linear one by introducing a slack variable, and solve the resulting semidefinite optimization problem using the numerical optimization package SeDuMi~\cite{Sturm99}.

\emph{Fidelity measures.}
The standard metric~\cite{Gilchrist05} for characterizing the similarity of a quantum process $\mathcal{E}$ to a reference process $\mathcal{F}$ can be calculated from their $\chi$-matrix representations as
\begin{equation}
F_\txt{pro}(\mathcal{E}, \mathcal{F}) = (1/d) \tr \sqrt{ \chi_E^{1/2} \chi_F \chi_E^{1/2} }.
\end{equation}
This quantity, termed \emph{process fidelity}, can also be used to characterize non-unitary channels. One finds $F_\txt{pro} (\mathcal{F}, \mathcal{F}) = 1$ for all trace-preserving channels $\mathcal{F}$. For non-trace-preserving channels, such as data qubit evolution conditioned on parity measurement, we use $d\chi/\tr(\chi)$ as the effective process matrix when calculating fidelities~\cite{Kiesel05}. For a unitary reference channel $\mathcal{F}$, $F_\txt{pro}$ can be calculated also from the $\mathcal{R}$ matrix representation as~\cite{Chow12}
\begin{equation}
F_\txt{pro}(\mathcal{E}, \mathcal{F}) = \tr\left(\mathcal{R}_E^\top \mathcal{R}_F\right)/d.
\end{equation}
Another widely-used process fidelity metric is the \emph{average gate fidelity} $F_\txt{ave}$, which is a measure of the average output state overlap between the experimental and reference processes. When the reference process is unitary, there is a linear mapping between $F_\txt{ave}$ as $F_\txt{pro}$~\cite{Gilchrist05}, namely
\begin{equation}
F_\txt{ave} = \frac{d F_\txt{pro} + 1}{d + 1}.
\end{equation}
\bibliography{../../../TeX/References/References_cQED}